\documentclass[12pt]{article}
\usepackage{amsfonts}
\usepackage{amssymb}
\usepackage{graphicx}
\usepackage{amsmath,graphicx,amsfonts}
\usepackage{epsfig}
\usepackage{subfigure}
\usepackage{dcolumn}

\def\siz{\small}

\def\be{\begin{equation}}
\def\ee{\end{equation}}
\title{General relativistic, nonstandard model for the dark sector of the Universe.}
\author{P.C. Stichel$^{1)}$  and W.J.
Zakrzewski$^{2)}$
\\
\siz
$^{1)}$Fakult\"at f\"ur Physik, Universit\"at Bielefeld, D-33501 Bielefeld, Germany, \\ \siz
e-mail:peter@physik.uni-bielefeld.de
\\ \\ \siz
$^{2)}$Department of Mathematical Sciences, University of Durham, \\
\siz Durham DH1 3LE, UK \\ \siz
 e-mail: W.J.Zakrzewski@durham.ac.uk
 }

\begin{document}
\date{}
\maketitle

\begin{abstract}
We present a general relativistic version of the self-gravitating fluid model for the dark sector of the Universe (darkon fluid) introduced in Phys. Rev. 80 (2009) 083513 and extended and reviewed in Entropy (2013) 559. This model contains no free parameters in its Lagrangian. The resulting energy-momentum tensor is dustlike with a nontrivial energy-flow. In an approximation valid at sub-Hubble scales we find that the present-day cosmic acceleration is not attributed to any kind of negative pressure but it is due to a dynamically determined negative energy density. This property turns out to be equivalent to a time-dependent spatial curvature. The obtained cosmological equations, 
at sub-Hubble scales, agree with those of the nonrelativistic model but they are given a new physical interpretation. Furthermore we have derived the self-consistent equation to be satisfied by the nonrelativistic gravitational potential produced by a galactic halo in our model from a weak field limit of a generalized Tolman-Oppenheimer-Volkoff equation.
\end{abstract}

\eject

 \section{Introduction} 

It is now pretty clear that the present Universe undergoes a phase of accelerated expansion (see the recent reviews \cite{ronea}, \cite{rtwoa}). On the other hand there exists an overwhelming evidence for the existence of gravitational effects on all cosmological scales (termed ``dark matter'') which cannot be explained by the gravitation of standard matter in the framework of General Relativity (GR) (see the review \cite{rthreea}). All of these data are in good agreement with a $\Lambda$-cold dark matter (CDM) cosmology (see \cite{ronea}, \cite{rtwoa} and the literature cited therein). But this $\Lambda$CDM model suffers, at least, from the following insufficiencies:
\begin{itemize}
\item Interpreted as the energy density of the vacuum the experimental value of $\Lambda$ turns out to be  too small by a factor of $10^{54}$ (see \cite{rfoura}).
\item None of the proposed DM-constituents has been observed (cp. \cite{rfivea})
\item There is a CDM-controversy on small scales \cite{rsixa}.
\end{itemize}
Other observations which are in disagreement with the $\Lambda$CDM model have been recently listed by Kroupa \cite{rsevena}.

One can find in the literature a large number of papers explaining either the accelerated expansion and/or dark matter by changing either the geometrical part of Einstein's field equations (termed modified gravity) or the matter part (addition of some scalar and/or tensor fields). We will not comment on either of these attempts (for details see {\it e.g.} the reviews \cite{ronea}, \cite{rtwoa} and the literature cited therein). But we want to point out that all these proposals are of a phenomenological nature, they contain either some new parameters or even free functions.
To overcome this freedom we need some new ({\it i.e.} unconventional) physics which, however, should be based on known physical principles ({\it e.g.} symmetry). Such a model containing no new constants in its Lagrangian and based on Galilean symmetry (minimal gravitational coupling of massless Galilean particles in agreement with the equivalence principle) has been presented in \cite{rone}, further developed in \cite{rthree} and reviewed and extended in \cite{rtwo}. This nonrelativistic, unified model for the dark sector of the Universe is an exotic fluid model,  termed darkon fluid model, which contains beside the standard hydrodynamic fields also a new vector field $\vec q(\vec x,t)$. This model describes successfully observational data for the transition from a decelerating to an acceleration phase of the Universe as well as the flat behaviour of galactic rotation curves \cite{rone}, \cite{rtwo}. 

The aim of the present paper is to present a general relativistic version of this model and to relate some approximate solutions of it to the corresponding solutions of its nonrelativistic counterpart.

The paper is organized as follows: To get a self-consistent paper and to have an appropriate starting point for its relativistic generalization we present in section 2 a short review of the nonrelativistic model \cite{rone}-\cite{rtwo}. In Section 3 we treat first the special relativistic generalization of the free model, discuss the different options to consider classical spin contributions and, after a Belinfante transformation, we introduce gravitation by the principle of minimal coupling. Also we discuss there the energy conditions. In Section 4 we consider the dynamics of the coupled system of the Einstein field equations and the relativistic darkon fluid equations of motion in spherical geometry.  Solutions of these equations at sub-Hubble scales which agree with the cosmological solutions obtained by the nonrelativistic model are treated in section 5. We show that these cosmological solutions turn out to be completely different from those of the FLRW model.  In section 6 we treat the same coupled system of equations in non-comoving coordinates and derive the self-consistent equation for the halo-gravitational potential derived in \cite{rtwo}, as a weak-field limit of the Tolman-Oppenheimer-Volkoff equation. Some final remarks are presented in section 7.

\section{Nonrelativistic, self-gravitating darkon fluid}

In \cite{rone} we have introduced nonrelativistic massless `particles' as a dynamical realization of the unextended Galilei group. These `particles' move in an enlarged twelve-dimensional phase space \cite{rtwo} consisting of
\begin{itemize}
\item The `particle' trajectory $\vec x(t)$
\item the momentum   $\vec p(t)$, canonically conjugate to $\vec x(t)$
\item the velocity vector $\vec y(t)$
\end{itemize}       
and
\begin{itemize}
\item the reduced boost vector  $\vec q(t)$  (called `pseudo-coordinate'), canonically
 conjugate to $\vec y(t)$.             .
\end{itemize}

In accordance with the Galilean algebra the corresponding `one-particle' Hamiltonian $H$ is given by
   \be
\label{eone}
H_0\,=\,p_iy_i,
\ee                                                                                                                                                                   
corresponding to, by a Legendre transformation, the Lagrangian
  \be
L_0\,=\,p_i(\dot x_i-y_i)\,+\,q_i\dot y_i
\label{etwo}
\ee                                                                                                                                                                     
and so giving the equations of motion (EOMs):
 \be
\label{ethree}                                                                                                                                                                       \dot x_i\,=\,y_i,\quad \dot p_i\,=\,0,\quad \dot q_i\,=\,-p_i,\quad \dot y_i\,=\,0.
\ee

But such a `particle' is not a classical particle in the usual sense as it is not detectable by any finite-sized
 macroscopic measurement device because 
\begin{itemize}
\item momentum and velocity vector are independent of each other and we have no ability to measure the momentum,
\item the boost vector $\vec q$        has, for fixed position $\vec x$          and velocity  $\vec y$             an arbitrary {\it i.e.} un-determined length.
\end{itemize}
For these reasons we have called these `particles' darkons \cite{rthree}; as they exist only as elements of an exotic fluid whose self-gravitating version is a substitute for what is usually called `dark energy' and `dark matter'.  

To introduce the coupling to gravitation represented by the field strength  $g_i(\vec x,t)$                     we have to require, in agreement with Einstein's equivalence principle, the validity of Newton's law
\be
\ddot x_i(t)\,=\,g_i(\vec x(t),t),
\label{efour}
\ee
                                                                                                                                                                   which will be realized if we add to $L_0$  an interaction part (minimal coupling)
\be
\label{efive}
L_{int}\,=\,-q_ig_i.
\ee
                                                                                                                                                                       
An important property of our darkons is the appearance of a macroscopic spin: The conserved total angular momentum                is given by the sum of the usual orbital angular momentum and a 2nd term which we call, for convenience, spin \cite{rtwo} (see also Mathisson \cite{rfour})
   \be
J_i\,=\,\epsilon_{ikl}(x_kp_l+y_kq_l).
\label{efivea}
\ee                                                                                                                                                                     
Note that the two terms in (\ref{efive}) act separately as
generators of rotations in the $\{ \vec x , \vec p \}$ resp. $\{ \vec y , \vec q \}$ parts of the phase space.


Now the question arises whether we can find some a simple physical system which mimics one darkon coupled to an external gravitational field.  To get this we start with the 2nd order Lagrangian 
 \be
L\,=\,-\dot q_i\dot x_i,
\label{B7}
\ee                                                                                                                                                            
which has been obtained from the Lagrangian (\ref{etwo}) by the elimination of the momentum    $p_i$. 
 By introducing the point transformation
\be
\label{B8}
(x_i,q_i)\rightarrow (x_i^+,x_i^-),\quad \hbox{with}\quad x_i^{\pm}\equiv x_i\,\pm\,\frac{q_i}{2m_0},
\ee                                                                                                                                               
where $m_0$       is a free mass-parameter introduced for dimensional reasons, we obtain
 \be
\label{B9}
L\,=\,\frac{m_0}{2}\left((x_i^-)^2-(x_i^+)^2\right).
\ee   
                                                                                                                                                           
This is a system of two non-interacting point particles with their masses having opposite sign but equal magnitude. The coordinates $x_i$     describe the motion of the geometric centre and $q_i/m_0$             describe the relative motion of the two particles (therefore we have called  $q_i$      `pseudo-coordinate')
 So, the vanishing mass of free darkons comes about by the cancellation of two mass terms. 

To arrive at the interaction Lagrangian (\ref{efive}) we start with the standard expression for the interaction of two massive particles with an external gravitational potential   
\be
\label{B10}
L_{int}\,=\,-m_0\,\phi(\vec x^{\,-})\,+\,m_0\,\phi(\vec x^{\,+}).
\ee
                                                                                                                                                                
If we insert the expressions for  $x_i^{\pm}$     from (\ref{B8}) and perform a Taylor expansion about  $x_i$        we obtain, in lowest order,                                                                                                                                                  
\be
\label{B11}
L_{int}\,=\,q_i\,\partial_i\phi\,+...,
\ee
which agrees exactly with our ansatz (\ref{efive}) if the field strength $g_i$ is given, as usual, by the gradient of a potential  $g_i=-\partial_i \phi$. 

                                  .
We need to add a word of caution: The description just given cannot be understood as a derivation of the Lagrangian (\ref{efive}) because the length of the vector  $q_i$          is unbounded and  it is by not small when compared to the length of the coordinate vector  $x_i$.  So the physical picture given above serves only for illustrative purposes.


To promote the `one-particle' picture to a self-gravitating fluid we replace the `one-particle' phase space coordinates 
$A_i=\{x_i,p_i,q_i,y_i\}$            by  the continuum labeled by  $\vec \xi \in R^3$              (comoving coordinates) $A_i(t)\rightarrow A_i(\vec \xi,t)$.

The Lagrangian for our darkon fluid then becomes
                                                                                                                                                                         
\be
L\,=\,\int \,d^3\xi\,[p_i(\dot x_i-y_i)\,+\,q_i(\dot y_i-g_i)]\,+\, L_{field}
\label{eseven}
\ee
where, as usual
\be
\label{eeight}
L_{field}\,=\,-\frac{1}{8\pi G}\,\int\,d^3x\,g_i^2(\vec x,t).
\ee
                                                                                                                                                                          
 The Lagrangian (\ref{eseven}) is invariant w.r.t. infinitesimal relabeling transformations $\vec \xi\rightarrow \vec \xi+\vec\alpha(\vec \xi)$                      
with   $\vec \nabla_{\xi}\cdot \vec \alpha=0$ leading to the conservation law \cite{rthree}, \cite{rtwo}
\be 
\label{enine}
\dot \theta_i\,=\,0\quad                  \hbox{where}\quad 
\theta_i\,\equiv\,-\frac{\partial \dot x_k}{\partial \xi_i}q_k\,+\,\frac{\partial x_k}{\partial \xi_i}\dot q_k
\ee
  which, after elimination of the momentum field $p_i$, allows us to reduce the EOM for $q_i$ to a 1-st order equation \cite{rthree}, \cite{rtwo}.                                                                                                                     
Then by means of the usual transformations from comoving coordinates    $\vec \xi$       to the fixed ones $\vec x=\vec x(\vec \xi,t)$           
we obtain from the Lagrangian formulation (\ref{eseven}) the Eulerian formulation given by the Lagrangian \cite{rtwo}
\be
\label{eten}
L=\int d^3x\,[nq_i(D_tu_i-g_i)\,-\, \theta(\dot n+\partial_k(nu_k))\,+\, n\alpha D_t\beta]\,+L_{field},\ee                                                                                                                                                      where $n(x,t)$ denotes the `particle' density. We have introduced the auxiliary field
\be
\nonumber
\theta_i(\vec x,t)\,\equiv\, \frac{\partial \xi_k}{\partial x_i}\,\theta_k(\vec \xi)\vert_{\vec \xi=\vec \xi(\vec x,t)}
\ee
  and its Clebsch-parameterization  $\theta_i\,=\,\partial_i\theta\,+\,\alpha\partial_i\beta$.
   
Furthermore  $\vec u$   denotes the velocity field 
$u_k(\vec x,t)\equiv \dot x_k(\vec \xi,t)\vert_{\vec \xi=\vec \xi(\vec x,t)}$                                                                      and $D_t$
the convective derivative $D_t\equiv \frac{\partial}{\partial t}+u_k\frac{\partial}{\partial x_k}.$ 

Note that the Hamiltonian corresponding to the Lagrangian (\ref{eten}) is not bounded from below. In \cite{rone} we have argued that this does not lead to any stability problems.  
                                                  
The equations of motion (EOMs) following from the Lagrangian (\ref{eten}) have been solved for
\begin{itemize}
\item  the isotropic, homogeneous case (cosmology) in \cite{rone} resp. \cite{rtwo} (see also section 5 of this paper),
\item the spherically symmetric, steady state case modeling halos \cite{rtwo} (see also section 6 of this paper).
\end{itemize}

\section{General relativistic approach}

\subsection{ Nongravitating, special relativistic case}
We start our discussion with the nongravitating {\it i.e.} special relativistic case for two reasons:
\begin{itemize}
\item to discuss the notion of (zero) rest-mass in our enlarged phase space,
\item to discuss the role of the spin-term within the energy-momentum tensor (EMT).
\end{itemize}
The relativistic generalization of the action corresponding to the free matter part of the Lagrangian (\ref{eten}) is then given by (we use the Minkowski metric $\eta_{\mu\nu}=\hbox{diag}(-1,+1,+1,+1)$)
\be
S\,=\,\int\,d^4x\,(nq_{\nu}Du^{\nu}\,-\,\theta\partial_{\nu}(nu^{\nu})\,+\,n\alpha D\beta),\label{eeleven}\ee                                                                                                                                                                                                                                                             
where we have defined the relativistic version of the convective derivative by $D\equiv u^{\lambda}\partial_{\lambda}$.
We also require that the velocity field $u^{\nu}$ obeys the usual constraint $u_{\nu}u^{\nu}=-1.$

From the Lagrangian (\ref{eeleven}) we derive the Euler-Lagrange EOMs
\be \partial_{\nu}(nu^{\nu})\,=\,0, 
\quad Du^{\nu}\,=\,0,\label{ethirteen}\ee
\be D\alpha\,=\,D\beta\,=\,D\theta\,=\,0,
\quad Dq_{\lambda}\,=\,q_{\nu}\partial_{\lambda}u^{\nu}\,+\,\theta_{\lambda}\label{efifteen}\ee                                                  
with 
\be
\label{esixteen}
\theta_{\lambda}\equiv \partial_{\lambda}\theta\,+\,\alpha \partial_{\lambda}\beta.
\ee
Then the fist part of the EOMs (\ref{efifteen}) is equivalent to the
EOM
\be
\label{B15}
D\theta_{\lambda}\,+\,\theta_{\nu}\partial_{\lambda}u^{\nu}\,=\,0
\ee
with the constraint
\be
\label{B16}
u^{\lambda}\theta_{\lambda}\,=\,0.
\ee

It is easy to see that 
\begin{itemize}
\item the four-momentum vector field, defined analogously to  the EOM $p_i=-\dot q_i$  in (\ref{ethree}) by 
$p_{\mu}\equiv -Dq_{\mu}$, is space-like (from the EOMs one deduces that $p^{\lambda}u_{\lambda}=0$),
\item the second EOM in (\ref{efifteen}) is invariant w.r.t. the gauge transformation $q_{\lambda}\rightarrow q_{\lambda}+\epsilon u_{\lambda}$,    {\it i.e.} we can fix the gauge by choosing   $q_{\lambda}u^{\lambda}=0$                             so that  $q_{\lambda}$         becomes space-like. 
\item the EOMs are invariant w.r.t. the shift symmetry $q_{\lambda}\rightarrow q_{\lambda}+c_{\lambda}$,
$\theta\rightarrow \theta-c_{\lambda}u^{\lambda}$, where $c_{\lambda}$ is a constant vector field. Note that this kind of shift symmetry is characteristic for Galileon theories (cp section 2.1 in \cite{rextra1}).
\end{itemize}

The fact that  $p_{\mu}$                    is a space-like vector field could easily lead to the wrong conclusion, that our darkons are tachyons (cp. appendix B in \cite{rone}). But, as argued by Weyssenhoff and Raabe \cite{rfive} in a similar context, we should define the rest-mass as the energy in the rest system of the `particle' given by $m_{0}=-u^{\lambda}p_{\lambda}$,                                      which, however, vanishes in our case.

The Poincare invariance leads to the existence of two conserved currents (cp. Appendix A in \cite{rsix})
\begin{itemize}
\item From translational invariance we get the canonical, nonsymmetric energy-momentum tensor (EMT)
\be \hat T^{\mu\nu}\,=\,np^{\mu}u^{\nu}
\label{eseventeen}
\ee                                                                                                                                                \item from the Lorentz invariance we get
\be
\label{eeighteen}
J^{\nu,\alpha\beta}\,=\,x^{\alpha}\hat T^{\beta\nu}\,-\,x^{\beta}\hat T^{\alpha \nu}\,+\,S^{\nu,\alpha\beta},\ee                                                                                                                                                where the spin tensor is given in our case by
\be
\label{enineteen}
S^{\nu,\alpha\beta}\,=\,nu^{\nu}(u^{\alpha}q^{\beta}-u^{\beta}q^{\alpha})
\ee
\end{itemize}
                                                                                                                                                                     
The conservation law  $\partial_{\nu}\hat T^{\mu\nu}=0$                     follows immediately from the EOMs. Furthermore, the EOMs also give us
\be
\label{etwenty}
\partial_{\nu}S^{\nu,\alpha\beta}\,=\,\hat T^{\alpha\beta}-\hat T^{\beta\alpha}
\ee
and so yield
\be
\nonumber \partial_{\nu}J^{\nu,\alpha\beta}\,=\,0.\ee


Note that the relativistic fluid described by the action (\ref{eeleven}) is a spin fluid which, as usually, is  described by a scalar density $n$, a four velocity $u^{\mu}$         and an anti-symmetric spin tensor $S^{\alpha\beta}$        defined in our case by (see eq. (\ref{enineteen}))
\be
\label{B19}
S^{\alpha\beta}\,=\,n(u^{\alpha}q^{\beta}-u^{\beta}q^{\alpha}).
\ee

Contrary to the standard relativistic spin fluid, our spin tensor (\ref{B19}) does not obey the Frenkel condition \cite{rseven}

 \be
\label{B20}
S^{\alpha\beta}u_{\beta}\,=\,0,
\ee
nor any other spin supplementary condition (for an exhaustive discussion of all these conditions see {\it e.g.} \cite{A16}). It is an important property of our fluid that we do not need such a supplementary condition   as the dynamics of the spin tensor (\ref{B19}) is completely fixed by the EOMs for $n$, $u^{\mu}$ and the field  $q^{\mu}$. 
    But for a standard relativistic fluid we need, besides the conservation law  $\partial_{\nu}T^{\mu\nu}\,=\,0$, also three additional equations to obtain a  well determined system. These additional equations are just the spin supplementary conditions ({\it e.g.} the Frenkel condition (\ref{B20})).



         \subsection{General relativistic dynamics}
According to Hehl \cite{reight} we have now two possibilities for coupling our relativistic fluid to gravity
\begin{itemize}
\item To gauge away the spin tensor by a Belinfante transformation \cite{rnine}
  \be
\label{etwentyone}
\hat T^{\mu\nu}\rightarrow T^{\mu\nu}=\hat T^{\mu\nu}+\frac{1}{2}\partial_{\lambda}(S^{\mu,\nu\lambda}+S^{\lambda,\nu\mu}+ S^{\nu,\mu\lambda})
\ee                                                                                                                                                                                                                                                                                                                                                                                                                                                                 
leading to a conserved, symmetric EMT
 \be
\label{etwentytwo} 
T^{\mu\nu}\,=\,n(u^{\mu}p^{\nu}+u^{\nu}p^{\mu})\,+\,\partial_{\lambda}(nu^{\mu}u^{\nu}q^{\lambda})
\ee                                                                                                                                                     

              This EMT may then be used as a source term in Einstein's field equations after we have       performed the substitutions (\ref{etwentythree}) (see eq. (\ref{ethirty})).

\item Consider spin as a dynamical variable by relating the spin tensor to the torsion tensor in the framework of a Riemann-Cartan space-time and use the canonical EMT (\ref{eseventeen}) as the source term in Einstein's field equations.
\end{itemize}

In this paper we prefer to use the first possibility as in this case we can reproduce, at sub-Hubble scales, the cosmological equations which are valid for the nonrelativistic darkon fluid (section 5).
To realize this we have to apply the principle of minimal gravitational coupling (cp. \cite{rten}):
So we perform the substitutions
 \be
\label{etwentythree} 
\eta_{\mu\nu}\rightarrow g_{\mu\nu}\quad                                                                             \hbox{and} \quad \partial_{\lambda}\rightarrow \nabla_{\lambda}
\ee                                                                           
in the special relativistic action (\ref{eeleven}). Here   $g_{\mu\nu}$                   is the metric tensor and $\nabla_{\lambda}$        is the covariant derivative $\nabla_{\lambda}A^{\nu}=\partial_{\lambda}A^{\nu}+\Gamma^{\nu}_{\lambda\sigma}A^{\sigma}$                                                                        where the elements of the connection $\Gamma^{\nu}_{\lambda\sigma}$           are given by the Christoffel symbols.

To obtain also Einstein's field equations from the principle of least action we have to consider the total action 
\be
\label{extra}
S\,=\,\int\, d^4x\,\sqrt{-g}(nq_{\nu}u^{\lambda}\nabla_{\lambda}u^{\nu}\,-\,\theta \nabla_{\nu}(nu^{\nu})\,+\,
n\alpha u^{\lambda}\partial_{\lambda}\beta)\,+\,S_{EH},
\ee
with the Einstein-Hilbert action $S_{EH}$ given by the well-known expression
               \be\label{etwentythreea}     S_{EH}\, =\,\frac{1}{16\pi G}\,\int\,d^4x\,\sqrt{-g}\,R, \ee                                                                                                               
where $g$ is the determinant of        $g_{\mu\nu} $, $R$ is the Ricci scalar, $n$ is the particle density and     $u^{\lambda} $  resp.   $q^{\lambda} $    are the velocity field resp. the relativistic generalization of the pseudo-coordinate field. The scalar fields       $\theta$,          $\alpha$ and   $\beta$  are Lagrange-multiplier fields which originate from the relabelling symmetry (see the nonrelativistic Lagrangian formulation in section 2). 

From the action (\ref{extra}) we derive the darkon fluid EOMs (which, alternatively, may be obtained by applying the substitution rule (\ref{etwentythree}) to the special-relativistic EOMs (\ref{ethirteen}-\ref{B16}))
\be
\label{etwentyfour} \nabla_{\lambda}(nu^{\lambda})=0,\qquad u^{\lambda}\nabla_{\lambda}u^{\nu}\,=\,0
\ee
\be
\label{etwentysix}
u^{\lambda}\nabla_{\lambda}q_{\nu}\,=\,q_{\lambda}\nabla_{\nu}u^{\lambda}\,+\,\theta_{\nu}\ee
and
\be
\label{etwentyseven}
 u^{\lambda}\nabla_{\lambda}\theta_{\nu}\,+\,\theta_{\lambda}\nabla_{\nu}u^{\lambda}\,=\,0\quad \hbox{with}\quad u^{\nu}\theta_{\nu}\,=\,0
\ee                                                                                                                                                       
where by (\ref{esixteen})  $\theta_{\nu}\equiv \partial_{\nu}\theta+\alpha\partial_{\nu}\beta$ and 
        Einstein's field equations (EFEs) are the standard ones
\be
\label{etwentyeight}
G^{\mu\nu}\,\equiv\,R^{\mu\nu}\,-\,\frac{1}{2}g^{\mu\nu}R\,=\,8\pi G T^{\mu\nu}.
\ee                                                                                                                                                                   
Here  $R^{\mu\nu}$ is the Ricci tensor and the EMT $T^{\mu\nu}$        is given by (\ref{ethirty}) given below.

We find again that the fields
        $q_{\lambda}$           and $p_{\lambda}\equiv-u^{\nu}\nabla_{\nu}q_{\lambda}$             obey the constraints
\be
\label{etwentynine}
u^{\lambda}q_{\lambda}\,=\,u^{\lambda}p_{\lambda}\,=\,0
\ee
and so they are space-like (recall that      $u^{\lambda} $   is time-like, normalized by         $u^{\lambda}u_{\lambda}=-1$).

\subsection{Energy-momentum tensor (EMT)}
The EMT (\ref{etwentytwo}), after having performed the substitutions (\ref{etwentythree}), becomes
 \be
T^{\mu\nu}\,=\,-n\,[u^{\mu}u^{\lambda}\nabla_{\lambda}q^{\nu}+(\mu\leftrightarrow \nu)]\,+\,\nabla_{\lambda}(nu^{\mu}u^{\nu}q^{\lambda}).
\label{ethirty}\ee                                                                                                                                                                         

By using the darkon fluid EOMs the expression (\ref{ethirty}) can be brought into its canonical form (see \cite{releven})
\be
\label{ethirtyone}
T^{\mu\nu}\,=\,\rho u^{\mu}u^{\nu}\,+\,k^{\mu}u^{\nu}\,+\,k^{\nu}u^{\mu},\ee
                                                                                                                                                                             where  for our model                                                    
\be                                                                                                \label{ethirtytwo}\rho=\nabla_{\lambda}(nq^{\lambda})\quad \hbox{and}\quad 
k^{\mu}\,=\,n\left(q^{\lambda}(\partial_{\lambda}u^{\mu}-\partial^{\mu}u_{\lambda})-\theta^{\mu}\right)
\ee
are the energy density  resp. the energy flow vector seen by an observer comoving  with the darkon fluid.

Usually the vector $k^{\mu} $        is called the `heat-flow vector'. But such a terminology assumes, at least implicitly, that we have a description of     $k^{\mu} $ and     $\rho$    in terms of a relativistic, irreversible thermodynamics (for the general framework see \cite{rnewa}, for an application to cosmology see \cite{rnewb}). But     $\rho$  and       $k^{\mu} $    are completely fixed in our case by the darkon fluid EOMs. So it is an open question whether they are accessible to a thermodynamic description or whether the arising energy flow is due to the generation of gravitational radiation. It is outside the scope of the present paper to consider this question.

Note that the expression (\ref{ethirtytwo}) for the energy density  $\rho$        is not positive definite! So at least the weak energy condition is violated. But, as will be shown in section 5, exactly this property of our model is crucial for the model's explanation of the present-day accelerated expansion of the Universe.
Energy conditions are constraints on the EMT of a general relativistic fluid which, originally, has been thought as being necessary for the fluid `to be physically reasonable' (see \cite{rtwelve} and the literature cited within).  But it is well known that {\it e.g.} the introduction of `dark energy' within the FLRW model (negative pressure with $\rho+3p<0$) violates the strong energy condition. This is in agreement with a very recent and general discussion in the framework of extended theories of gravitation  \cite{rthirteen}, which comes to the conclusion that the violation of energy conditions is a general property in the presence of dark energy.

 \section{Dynamics in spherically symmetric geometry}
The non-accelerated fluid (geodesic) motion (\ref{etwentyfour}) allows the consideration of synchronous comoving ($u^{\mu}=\delta^{\mu}_0)$,
spherically symmetric coordinates  defined by the metric
\be
\label{ethirtythree}
ds^2\,=\,-dt^2\,+\, B^2(t,r)dr^2\,+\,Y^2(t,r) d\Omega^2.
\ee                                                                                                                                                                         
For this metric the space-like vectors   $q_{\mu}$  and   $\theta_{\mu}$ have only a non-vanishing radial component   
\be
\label{ethirtyfour}
q_{\mu}=qs_{\mu},\quad \theta_{\mu}=\tilde \theta s_{\mu}\qquad  \hbox{with}\qquad s_{\mu}\equiv (0,B),
\ee 
 where here, and in the following, the first component of a 2-dim vector describes the time-component and the 2nd one the radial component.   
                                            
The darkon fluid equations (\ref{etwentyfour}), (\ref{etwentysix}) and (\ref{etwentyseven}) have the following form resp. solutions
\be 
\label{ethirtyfive}
 n(t,r)=\frac{n_0(r)}{BY^2},\quad \left(\frac{q}{B}\right)^.=\frac{\alpha(r)}{B^2}, \quad \tilde \theta(t,r)=\frac{\alpha(r)}{B},\ee                                                                                                                                                  
where  $n_0(r)$  resp.   $\alpha(r)$  are arbitrary integration functions. The energy density  $\rho$        defined by (\ref{ethirtytwo}), then becomes,   in terms of $q$ and the metric 
 \be
\label{ethirtysix}  
\rho\,=\,\frac{1}{BY^2}\left(\frac{n_0q}{B}\right)^{'}
\ee                                                                                                                                                                          
and obeys, due to the 2nd eq. in (\ref{ethirtyfive}), the local energy conservation equation
\be
\label{ethirtyseven}
\dot \rho\,+\,\rho \left(\frac{\dot B}{B}+2\frac{\dot Y}{Y}\right) \,-\,\frac{1}{BY^2}\left(\frac{\alpha(r)n_0(r)}{B^2}\right)^{'}\,=\,0,
\ee
where and in what follows $^{'}$ denotes the derivative w.r.t. $r$.
                                                                                                 
Note that the velocity field  $u^{\mu}$           has vanishing vorticity in the spherically symmetric case. Therefore the energy flow vector  $k^{\mu}$         reduces to  
\be 
\label{ethirtyeight}
k^{\mu}=-n\theta^{\mu}.\ee                                                                                           
The Einstein-field equations (\ref{etwentyeight}) now become (cp. eq. (7) with $A=1$ in \cite{rfourteen}) 
\be
\label{ethirtynine}
2\frac{\dot B}{B}\frac{\dot Y}{Y}\,+\,\frac{1+\dot Y^2}{Y^2}\,-\,\frac{Y^{'2}}{Y^2B^2}\,-\,\frac{2}{YB}\left(\frac{Y^{'}}{B}\right)^{'}\,-\,\frac{\kappa}{BY^2}\left(\frac{n_0 q}{B}\right)^{'}\,=\,0,\ee
\be 
\label{efourty}
2\frac{\ddot Y}{Y}\,+\, \frac{1+\dot Y^2}{Y^2}\,-\,\frac{Y^{'2}}{Y^2B^2}\,=\,0
\ee
\be
\label{efourtyone}
\frac{\ddot B}{B}\,+\,\frac{\ddot Y}{Y} \,+\,\frac{\dot B\dot Y}{BY}\,-\, \frac{1}{BY}\left(\frac{Y^{'}}{B}\right)^{'}\,=\,0,
\ee                                                                                                                                                                              
\be
\label{efourtytwo}
-\frac{\kappa}{2}n_0(r)\alpha(r)\,=\,YB^2\left(\frac{Y^{'}}{B}\right)^{.}
\ee
with $\kappa\equiv 8\pi G$
where (\ref{ethirtynine}), (\ref{efourty}), (\ref{efourtyone}) and (\ref{efourtytwo}) represent, respectively, the $00$, $rr$, tangential and $0r$-components of (\ref{etwentyeight}).


As a consequence of the covariant conservation of the Einstein tensor
\be
\label{B42}
\nabla_{\nu}G^{\mu\nu}\,=\,0
\ee                                                                                                                                                                 
the four EFEs (\ref{ethirtynine}-\ref{efourtytwo}) are not independent of each other. Explicit calculations lead to the following results:
\begin{itemize}
\item The third EFE (\ref{efourtyone}) is a consequence of the 2nd and the 4th EFEs (\ref{efourty}) and (\ref{efourtytwo}).
\item  The time derivative of  $BY^2\times$ l.h.s. (\ref{ethirtynine}) vanishes as a consequence of the other EFEs and the second EOM in (\ref{ethirtyfive}).
\end{itemize}
These dependencies give rise to consistency relations which have to be respected if we consider approximate solutions of the EFEs (see section 5).

 For the discussion of approximate cosmological equations (see section 5) it is also advantageous to express at least partially the EFEs in terms of some kinematic resp. geometric quantities. Kinematic quantities are defined by the EMT (see (\ref{ethirtyone}) with (\ref{ethirtyeight})) and the decomposition of the 4-velocity gradient [20]
\be
\label{B43}
\nabla_{\nu}u_{\mu}\,=\,\sigma_{\nu\mu}\,+\,\frac{1}{3}\hat \theta h_{\mu\nu},
\ee                                                                                                                                                                
where    $\hat \theta\equiv \nabla_{\nu}u^{\nu}$   is the volume expansion scalar and   $\sigma_{\nu\mu}$            is the traceless shear tensor which for our metric (\ref{ethirtythree}) takes the form 
\be
\label{B44}
\sigma_{\nu\mu}\,=\,\sqrt{3}\,\sigma\,(s_{\nu}s_{\mu}-\frac{1}{3}h_{\nu\mu})\ee                                                                                                                                                      with  $\sigma\equiv \frac{1}{\sqrt{3}}\left(\frac{\dot B}{B}-\frac{\dot Y}{Y}\right).$

The tensor  $h_{\nu\mu}\,=\,g_{\nu\mu}+u_{\nu}u_{\mu}$  projects onto the space orthogonal to the 4-velocity  $u^{\mu}$.
Note that in our case the decomposition (\ref{B43}) contains neither a vorticity nor an acceleration term.

As a geometric quantity we also introduce the spatial Ricci scalar $^3R$               (for its definition see \cite{A25}, section 1.3.5).

Next we consider the following equations which are derived from the EFEs (\ref{ethirtynine}-\ref{efourtytwo}) (see \cite{A25},\cite{A26}):
\begin{itemize}
\item The Raychaudhuri-Ehlers (RE) equation
\be
\label{B45}
\frac{\ddot a}{a}\,=\,-\frac{2}{3}\sigma^2\,-\,\frac{4\pi G}{3}\rho,
\ee
where the generalized scale factor $a(t,r)$is defined  by
\be
\label{B46} 
\frac{\dot a}{a}\equiv \frac{1}{3}\hat \theta.
\ee                                                                                                                                    \item The generalized Friedmann equation
\be
\label{B47}
\dot a^2\,+\,\frac{a^2}{6}\left(^3R-2\sigma^2\right)\,=\,\frac{8\pi G\rho a^2}{3},\ee                                                                                                                                                   or, if we define an effective spatial curvature     $K_{eff}$            by
\be 
\label{B48}                                                                                                                                                  
K_{eff}(t,r)\equiv \frac{a^2}{6}(^3R-2\sigma^2),
\ee
eq. (\ref{B47}) becomes
\be
\label{B49}
\dot a^2\,+\,K_{eff}\,=\,\frac{8\pi G \rho a^2}{3}.
\ee
Note that we have    $K_{eff}  = (0, \pm 1)$ in the FLRW case.
\item From these equations and the relation  
\be
\label{B50}
\nabla_{\nu}k^{\nu}\,=\,-\frac{1}{BY^2}\left(\frac{n_0 \alpha}{B^2}\right)'
\ee
               which follows from  (\ref{ethirtyeight}) and the 1st and 3rd EOM in (\ref{ethirtyfive}),
               we easily derive the time derivative of  $K_{eff}$ 
\be  
 \label{B51}
\dot K_{eff}\,=\,\frac{4}{9}\,a^2\hat \theta \sigma^2\,+\,\frac{8\pi G a^2}{3 BY^2}\left(\frac{n_0\alpha}{B^2}\right)'
.\ee                                                                                                                                         So we obtain the local energy conservation equation (\ref{ethirtyseven}) in the form
\be
\label{B52}
\dot \rho \,+\,3 \frac{\dot a}{a}\rho \, - \,\frac{3}{8\pi G a^2}\dot K_{eff}\,+\,\frac{\hat \theta \sigma^2}{6\pi G}\,=\,0.
\ee 
\end{itemize}                                                                                                                                                    

Note that these equations are not independent: The RE-eq. (\ref{B45}) follows by time differentiation of (\ref{B49}) and the use of (\ref{B51}) and (\ref{B52}).

What about the contribution of baryonic matter within our model?  Suppose we describe baryonic matter, averaged over small scale inhomogeneities, by dust moving with the same four velocity then the darkon fluid. Then our cosmological equations contain only the total energy density $\rho$  given by the sum of the darkon fluid and the baryonic dust contribution. We are unable to discriminate between both contributions to $\rho$. This can be seen as follows: 
To take into account baryonic matter we have to add to the EMT (\ref{ethirtyone}) a dust contribution   $T^B_{\mu\nu}$      ($\rho^B$  is the baryonic energy density)
 \be
\label{eseventynine} 
T^B_{\mu\nu}\,=\,\rho^Bu_{\mu}u_{\nu}
,\ee 
which is separately covariantly conserved
\be
\label{eeighty}
\nabla_{\rho}\,T_{\mu\nu}^{B}g^{\rho\mu}\,=\,0.
\ee                             

Then  $\rho^B$  obeys the local energy conservation equation
 \be
\label{eeightyone} 
\dot \rho^B\,+\,\rho^B\left(\frac{\dot B}{B}\,+\,2\frac{\dot Y}{Y}\right)\,=\,0
\ee                                                                                                                                                           
and we have to add  $\kappa \rho^B$  to the r.h.s. of the 1st Einstein-field eq. (\ref{ethirtynine}).
So only the total energy density appears in (\ref{ethirtyseven}). But the solution of the energy conservation eq. (\ref{ethirtyseven}) for the darkon fluid is only fixed modulo a solution of the corresponding homogeneous eq. which is just given by (\ref{eeightyone}).


Let us next show that, for isotropic coordinates,      
\be 
\label{efourtythree}
Y(t,r)\,=\,rB(t,r).
\ee
(\ref{efourty})-(\ref{efourtytwo}) enforces    $\alpha(r)=0$:
By equating (\ref{efourty}) and (\ref{efourtyone}) in the isotropic case (\ref{efourtythree}) we eliminate the time derivatives and obtain the well- known result  (see \cite{rnewc})
\be 
\label{efourtyfour}
\left(\frac{1}{r}\left(\frac{1}{B}\right)^{'}\right)^{'}\,=\,0
\ee 
 with the solution 
\be
\label{efourtyfive}
B(t,r)\,=\,\frac{a(t)}{1+\frac{r^2}{4}K(t)}
\ee                                                       
where $a(t)$ resp. $K(t)$ are arbitrary functions of $t$. Now inserting (\ref{efourtythree}), (\ref{efourtyfive}) into (\ref{efourty}) leads by a straightforward calculation to $K(t)= K =$ const and, therefore, to the
 vanishing r.h.s. of (\ref{efourtytwo}).
But then the EMT contains only a pure dust term which leads to a trivial cosmology (presence of only a decelerating phase). 

\section{Solutions at sub-Hubble scales}

Unfortunately we are unable to solve Einstein's eq.s for  $\alpha\ne 0$ exactly.
 So let us look for approximate solutions at sub-Hubble scales  $\frac{r}{r_0}=\epsilon\ll 1$  ($r_0$   = Hubble radius) and take correspondingly for the derivatives (cp. \cite{rfifteen})                                                        
 \be
\label{efourtysix}  
\partial_r\,=\,O(\epsilon^{-1})\quad                                                                \hbox{and}\quad \partial_t\,=\,O(\epsilon^{-\frac{1}{2}}).
\ee
                                                                                            
From (\ref{efourty}) we obtain the exact relation
\be 
\label{efourtyseven}
B\,=\,\frac{Y^{'}}{(1-b)^{\frac{1}{2}}}
\ee                                                         with 
\be
\label{efourtyeight}
b\,=\,-(2\ddot Y Y+\dot Y^{2})
\ee

Now we consider those metrics which have         $Y(t,r)\propto r$                   for small $r$. Then we have     $b=O(\epsilon) $ and therefore $b$ may be treated as a perturbation.

Next we obtain, in leading order, from (\ref{efourtytwo})
 \be
\label{efourtynine}
\dot b(t,r)\,=\,8\pi G \frac{n_0 \alpha(r)}{Y Y^{'2}},
\ee                                                                                                                                                                          
which, when compared with (\ref{efourtyeight}), leads to the consistency relation
\be
\label{efifty}
-\left(\ddot YY^2\right)^{.}\,=\,4\pi G \frac{n_0 \alpha}{Y^{'2}}.
\ee 
Then, in accordance with the interdependences of the EFEs described in section 4, the 3rd Einstein eq. (\ref{efourtyone}) is also fulfilled in leading order.

Instead of the 1st EFE (\ref{ethirtynine}), we use the RE eq. (\ref{B45}) yielding,
in leading order,
\be
\label{efiftyone}  
2\ddot Y Y^{'}\,+\,\ddot Y^{'}Y\,=\,-\frac{4\pi G}{Y}\left(\frac{n_0 q}{Y^{'}}\right)^{'}
\ee                                                                                                                                                                         
which, after multiplication by $Y$, can be integrated to give
\be 
\label{efiftytwo}
\ddot Y\,Y^2\,=\,-\,4\pi G\,\frac{n_0 q}{Y^{'}}\,+\,f(t),
\ee                                                                                                                                                                           
where $f(t)$  is an integration function. But if we differentiate (\ref{efiftytwo}) w.r.t. the time $t$ and use the EOM (\ref{ethirtyfive}) for $\frac{q}{B}$,  given in leading order by  
\be
\label{efiftythree}
\left(\frac{q}{Y^{'}}\right)^{.}\,=\,\frac{\alpha}{Y^{'2}}.\ee                                                                                                
 we obtain, by comparison with the consistency relation (\ref{efifty}), that $f$ must be a constant. To get an analytic solution we put $f$ equal to zero. 
So finally we have to solve the coupled system of equations (\ref{efiftytwo}) for $f=0$ and (\ref{efiftythree}). To do this we consider a separation ansatz for $Y$
\be
\label{efiftyfour}
Y(t,r)\,=\,a(t)y(r)
\ee                                                                                                                                                                    leading by (\ref{efiftytwo}) to a separable form for $q$ 
\be
\label{efiftyfive}
q(t,r)\,=\,q_0(t)q_1(r)
\ee                                                                                                  
where, due to (\ref{efiftythree}), we may normalize $q_1$  so that
 \be
\label{efiftysix}
q_1(r)\,=\,4\pi G\frac{\alpha(r)}{y'(r)}
\ee
Then we get for  $q_0$   the equation   
\be
\label{efiftyseven}
\left(\frac{q_0}{a}\right)^{.}\,=\,\frac{1}{4\pi G a^2}.\ee                                                                                                             
Finally from (\ref{efiftytwo}) we obtain
 \be
\label{efiftyeight}
\ddot a a^3\,=\,4\pi G K_1q_0
\ee     
\be
\label{efiftynine}
\hbox{where}\qquad y^3y^{'2}\,=\,-\frac{4\pi G}{K_1}\alpha n_0,
\ee  and                                             
 $K_1$  is an arbitrary constant. 

So the $r$-dependence of our solutions is completely specified by the choice of the integration functions $n_0(r)$               and $\alpha(r)$             (for the case of cosmology we refer to the next subsection).  
We note that the separable forms of   $Y$  and  $q$   lead also to a separable form for the energy density   $\rho$          (\ref{ethirtysix}). Therefore the appearance of a Perpetuum Mobile of the third kind (continuous transfer of energy from one space region to another one) as advocated by Ivanov \cite{rsixteen} is excluded.

\subsection{Connection with the nonrelativistic darkon fluid cosmology}
Note that (\ref{efiftyseven}) and (\ref{efiftyeight}) have exactly the form of the cosmological equations derived for the nonrelativistic darkon fluid in \cite{rone} resp. \cite{rtwo}.
 But which choice has to be made for the two free functions  $n_0(r)$  and   $\alpha(r)$?

For the cosmological solutions to be viable we have to require that the energy density  $\rho$ as well as the darkon density $n$ are functions of time only. Now taking  $\rho$   from (\ref{ethirtysix}) in leading order and using the ans\"atze (\ref{efiftyfour}), (\ref{efiftyfive}) together with (\ref{efiftynine}) we obtain 
 \be
\label{esixty}
\rho(t,r)\,=\,-\frac{K_1q_0(t)}{a^4(t)y'(r)y^2(r)}(y^3(r)y'(r))'\ee
So to get $\rho=\rho(t)$                                                                                                                                                                               we have to choose $y(r) = r\times$const. as expected. By fixing the scale for $r$ we can put this constant equal to one and we obtain
\be
\label{esixtyone}
\rho(t)\,=\,-\frac{3K_1q_0}{a^4}.
\ee                                                                                                                                                                             
Analogously, the requirement that $n = n(t)$ leads, due to the 1st eq. in (\ref{ethirtyfive}), to    
\be
\label{esixtytwo}
n_0(r)\,=\,r^2n_{00}
\ee                                 
and therefore, due to (\ref{efiftynine}) to  
\be
\label{esixtythree}
\alpha(r)\,=\,r\alpha_0
\ee                                                       with 
 \be
\label{esixtyfour}
-4\pi Gn_{00}\alpha_0\,=\,K_1,\ee
where $n_{00}$ and $\alpha_0$ are arbitrary constants.


To get for the cosmological equations (\ref{efiftyseven}), (\ref{efiftyeight}) exactly the form derived in \cite{rtwo} for the nonrelativistic darkon fluid model we define the function $g(a(t))$ by
 \be
\label{B64}
g(a)\,\equiv\,4\pi G\,K_1\frac{q_0}{a}
\ee                                                                                                                                                                  
Then the 1st eq. in (\ref{ethirtyfive}) as well as the eq.s (\ref{efiftyseven}) and (\ref{efiftyeight}) become
 \be
\label{B65}
n(t)\,=\,\frac{n_{00}}{a^3},\quad \dot g\,=\,\frac{K_1}{a^2} \quad \hbox{and}\quad \ddot a\,=\,\frac{g(a)}{a^2}.
\ee                                                                                                                                                                  
 The equations given in (\ref{B65}) are identical with eq.s (120)$-$(122) in \cite{rtwo}. 
 For the energy density (\ref{esixtyone}) we obtain 
 \be
\label{B66}
\rho(t)\,=\,-\frac{3g}{4\pi G a^3}.
\ee                                                                                                                                                                   
As shown in \cite{rtwo} the 2nd and 3rd eq. in (\ref{B65}) give rise to two conserved quantities $K_{2,3}$     
 \be
\label{B67}                                                                                                                                                                   K_2=\dot aK_1\,-\,\frac{1}{2}g^2
\ee
and 
\be
\label{B68}
K_3\,=\,\frac{g^3}{6}\,+\,K_2\,g\,+\,\frac{K_1^2}{a}.
\ee
                                                                                                                                                             
These conservation laws lead for the choice  $K_{2,3}>0$                 (and consequently   $K_1>0$) 
to a transition from an early decelerating phase of the Universe to an late accelerating phase with a transition redshift \cite{rtwo}
\be
\label{B69}
1\,+\,z_t\,=\,\frac{K_3}{K_1^2}.
\ee 
                                                                                                                                                                    
What have we achieved? 

We have shown that at sub-Hubble scales our general relativistic model agrees with our nonrelativistic model with the latter showing the observed transition from a decelerating to an accelerating phase of the Universe. But the observed transition redshift lies somewhere between $\frac{1}{2}$ and 1(cp. \cite{A28}) which, at least for the $\Lambda$CDM-model, corresponds to a luminosity distance of the order of the Hubble radius (cp. Fig. A 2.3 in \cite{A29}). So we are not yet able to prove this transition for our relativistic model. But what remains is an interesting result on the behaviour of the spatial curvatures at sub-Hubble scales. To get this we insert first of all our last results into (\ref{efourtyeight}) resp. (\ref{efourtynine}) and obtain
 \be
\label{B70}  b(t,r)\,=\,r^2\,K(t)\quad \hbox{with}\quad K(t)\,=\,-(2\ddot aa+a^2)
\ee                                                                                                                                                          
resp.                                                                                                                                              
 \be
\label{B71} 
\dot K\,=\,-\frac{2K_1}{a^3}.
\ee                                                                                                                                                                     
So the metric function $B(t, r)$ (\ref{efourtyseven}) becomes 
 \be
\label{B72} 
B^2(t,r)\,=\,\frac{a^2(t)}{1-r^2K(t)}.
\ee

Next we show that $ K(t)$  may be identified with the effective spatial curvature $K_{eff}$     introduced in (\ref{B48}).

First of all we insert the energy density  $\rho$      from (\ref{B66}) into the 3rd  eq. of (\ref{B65}) and obtain
the standard cosmological RE-eq. with vanishing pressure
 \be
\label{B73}
\frac{3\ddot a}{a}\,=\,-4\pi G \rho.
\ee
                                                                                                                                                                        
Eliminating $\ddot a$          in (\ref{B70}) by means of (\ref{B73}) we obtain the fundamental Friedmann eq. \cite{rseventeen}
\be
\label{B74}
\dot a^2\,+\,K(t)\,=\,\frac{8\pi G}{3}\rho a^2,
\ee  
but with a time dependent spatial curvature $K(t)$,                                                                                                                                                                  
which agrees with the generalized Friedmann eq. (\ref{efifty}) if we identify at sub-Hubble scales
 \be
\label{B75}  
K(t)\,=\,K_{eff}.
\ee    
                                                                                                                                                                    
For the sake of completeness we remark that the energy conservation eq. (\ref{ethirtyseven}) now takes the form
 \be
\label{B76}
\dot\rho\,+\,3\frac{\dot a}{a}\rho\,-\,\frac{3}{8\pi G a^2}\dot K\,=\,0,
\ee                                                                                                                                                                          
which is in agreement with its general form given in (\ref{B52}) (note that the shear $\sigma$    vanishes in the leading order). 

Finally we may express $K(t)$         in terms of the function $g(a)$ which has been defined in (\ref{B64}) and can be determined by the solution of the cubic eq. (\ref{B68})
\be
\label{B77}
K(a)\,=\,-\left(2\frac{g(a}{a}\,+\,\frac{1}{K_1^2}\,(K-2+\frac{1}{2}g(a)^2)^2\right).
\ee

Let us summarize: At least the observed present-day accelerated expansion of the Universe is determined in our general relativistic model by a negative energy density (see eq. (\ref{B73})) or, equivalently, by a time dependent spatial curvature (see eq. (\ref{B74})). But the behaviour of the spatial curvature at larger redshifts deserves for further studies.

Note that a time dependence of the spatial curvature with a possible sign change during evolution is already known for the Stephani solution of the EFEs \cite{reightteen}, \cite{rnineteen}.



\section{Non-comoving  coordinates and modeling of halos}
In this section we consider the darkon fluid moving in the radial direction relative to the cosmic rest system (CRS). The metric in the CRS is assumed to be given by Schwarzschild-like coordinates. 
We will
\begin{itemize}
\item derive the darkon fluid EOMs and the Einstein field equations by choosing the energy-frame (vanishing heat flux) for the CRS (see \cite{rtwenty}),
\item look for weak field solutions which arise to be equal to the nonrelativistic  stationary solutions derived in \cite{rtwo} modeling halos.
\end{itemize}
\subsection{Cosmic rest system (CRS)}
Schwarzschild-like coordinates are defined by the spherically symmetric metric
 \be
\label{eeightytwo} 
ds^2\,=\,-e^{2\phi(t,r)}dt^2\,+\,e^{2\lambda(t,r)}dr^2\,+\,r^2 d\Omega^2.
\ee                                                                                                                                                                      This metric is assumed to be valid in the CRS defined by a time-like unit vector $n^{\nu}$        and a space-like unit vector   $s^{\nu}$ \be
\label{eeightythree}
n^{\nu}\,\equiv\,(e^{-\phi},0),\qquad s^{\nu}\,\equiv\,(0,e^{-\phi})
\ee                                                                                                                                                             such that the CRS becomes the energy frame (vanishing heat flux) {\it i.e.} the EMT (\ref{ethirtyone}) takes in the CRS frame the form
\be
\label{eeightyfour}
T^{\mu\nu}\,=\,\rho^{\star}(t,r)n^{\mu}n^{\nu}\,+\,p_r^{\star}(t,r)s^{\mu}s^{\nu},
\ee    
where $\rho^{\star}$, resp. $p_r^{\star }$ are the energy density, resp. the radial pressure in the CRS.                                                                                                                                                         
Note that (\ref{eeightyfour}) contains no transversal pressure   $p_t$  as (\ref{ethirtyone}) is free of it (radial movement does not change  $p_t$). 

The darkon fluid is assumed to move with velocity $v$ in the radial direction relative to the CRS. Then the four-velocity   $u^{\mu}$        resp. the vector   $\theta^{\mu}$   are given by 
 \be
\label{eeightyfive}
u^{\mu}\,=\,\gamma(n^{\mu}+vs^{\mu}),\quad \theta^{\mu}\,=\,\tilde \theta \gamma(vn^{\mu}+s^{\mu}),\ee                                                                                                                                                                    
where  $\gamma(v)\equiv(1-v^2)^{-\frac{1}{2}}$.
                                            
Comparing (\ref{ethirtyone}) with (\ref{eeightyfour}) and  using (\ref{eeightyfive}) we obtain
  \be
\label{eeightysix}
\rho^{\star}\,=\,\frac{\rho}{1+v^2}\qquad                                                                                 \hbox{and}\qquad p_r^{\star}\,=\,-\rho^{\star}v^2,
\ee                                                                            
where  $v(t, r)$ is determined by the requirement of the vanishing heat flux in the CRS
 \be
\label{eeightyseven}
\rho^{\star}v\,-\,n\tilde\theta\,=\,0.
\ee                                                                                                                                                                     
\subsection{Einstein's field equations}
With the metric (\ref{eeightytwo}) and the EMT (\ref{eeightyfour}) we get for the Einstein-field equations (see \cite{rtwentyone} and the literature cited therein)
\be
\label{eeightyeight}
\kappa \rho^{\star}\,=\,\frac{1}{r^2}(r(1-e^{-2\lambda}))',\ee
\be
\label{eeightynine}
\kappa p_r^{\star}\,=\,\frac{1}{r^2}\left(-1\,+\,e^{-2\lambda}(1\,+\,2r\phi')\right),
\ee
\be
\label{eninety}
0\,=\,\phi''\,+\,\phi'^{2}\,-\,\phi'\lambda'\,+\,\frac{\phi'-\lambda'}{r},
\ee
in which we had already used (\ref{eninetyone}),
and 
\be
\label{eninetyone}
0\,=\,\dot \lambda.
\ee                                                                                                                                                                   
Using (\ref{eninetyone}) in (\ref{eeightyeight}) we get immediately  
\be
\label{eninetytwo}
\dot\rho^{\star}\,=\,0
\ee                                                                                                      
\subsection{ Darkon fluid EOMs}
From the darkon fluid EOMs (\ref{etwentyfour}-\ref{etwentyseven}) and (\ref{eninetyone}) we get by using the metric (\ref{eeightytwo}) and eq. (\ref{eeightyfive}) for $u^{\mu}$, resp. $\theta^{\mu}$:
\begin{itemize}
\item the continuity eq. (1st eq. in (\ref{etwentyfour})) becomes
 \be
\label{eninetythree}
0\,=\,e^{-\phi}\,\dot n\,+\,e^{-\lambda}\frac{(r^2nv)'}{r^2} 
\ee                                                                                                                                                                         
\item the Euler eq. (2nd eq. in  (\ref{etwentyfour}))  becomes (cp. \cite{rtwentyone}, eq. (17))
 \be
\label{eninetyfour}                                                                                                           
0\,=\,e^{-\phi}\,\dot v\gamma^2\,+\,e^{-\lambda}(vv'\gamma^2\,+\,\phi^{'})
\ee                                                                                                                                                                                                                                 
and
\item by using   $q^{\mu}=q\gamma(vn^{\mu}+s^{\mu})$   we obtain from (\ref{etwentysix}) 
\be
\label{eninetyfive}
e^{-\phi}\gamma\dot q\,+\,\gamma e^{-\lambda}(vq'-qv')\,=\,\tilde \theta,
\ee                                                                                                                                                                            
  where,  due to (\ref{etwentyseven}),  $\tilde \theta$   obeys the EOM 
\be
\label{eninetysix}
e^{\lambda}(\tilde \theta\gamma)^{.}\,+\,(\tilde \theta \gamma v e^{\phi})'\,=\,0.\ee
\end{itemize}                                                                                                                                                                            
                                                                                                                                                                          
 Finally we may express   $\rho^{\star}$ defined by
(\ref{ethirtytwo}) and (\ref{eeightysix})        in terms of the metric and the darkon fluid fields and we get
 \be
\label{eninetyseven}
\rho^{\star}\,=\,\frac{\gamma}{r^2}\,e^{-\lambda}\,(r^2 q n)'.
\ee                                                                                                                                                                            

Sometimes it is useful to use instead of the darkon fluid EOMs the EOMs for  $\rho^{\star}$          resp.  $p_r^{\star}$        which follow from the covariant conservation of the EMT  
\be
\label{eninetyeight} 
\nabla_{\mu}T^{\mu\nu}\,=\,0\ee  
We recall that (\ref{eninetyeight}) can be derived either from the Bianchi identities for the Riemann tensor or directly from the darkon fluid EOMs. From the time-like part of (\ref{eninetyeight}) we reproduce (\ref{eninetytwo}) whereas the space-like part leads to the generalized Tolman-Oppenheimer-Volkoff (TOV) equation (see \cite{rtwentytwo}) which in our case takes the form
 \be
\label{eninetynine}                                                                                                                                                                            
(\rho^{\star}\,+\,p_r^{\star})\phi'\,+\,2 \frac{p_r^{\star}}{r}\,+\,(p_r^{\star})'\,=\,0.
\ee

{\bf Elimination of  $\tilde\theta$, a conservation law}

By inserting     $\tilde\theta$       from (\ref{eeightyseven}) into
\begin{itemize}
\item (\ref{eninetysix}) and  using (\ref{eninetytwo}), (\ref{eninetyfour}) and (\ref{eninetynine}) we observe that (\ref{eninetysix}) is identically satisfied.
\item  (\ref{eninetyfive}) and using (\ref{eninetyseven}) for $\rho^{\star}$   we obtain
\end{itemize} 
\be
\label{esto}                                                                                                                                                                   e^{-\phi}\dot  q\,=\,e^{-\lambda}\frac{q}{nr^2}\,(vnr^2)'.
\ee
Combining (\ref{eninetythree}) with (\ref{esto}) leads to the conservation law
\be
\label{estoone}
  (nq)^{.}\,=\,0.
\ee

\subsection{Some exact relations}
Here we derive some exact expressions which follow from the coupled system of Einstein field eq.s and darkon fluid EOMs.
 
By using (\ref{eninetyone}), (\ref{eninetytwo}) and (\ref{estoone}) we conclude from (\ref{eninetyseven}) that   
\be
\label{estotwo}
\dot v \,=\,0
\ee                                                                 
and therefore the 2nd eq. in (\ref{eeightysix}) leads  to                                                                                                  
\be
\label{estothree}
\dot p_r^{\star}\,=\,0,
\ee
which, when used in the 2nd  Einstein eq. (\ref{eeightynine}) gives
\be
\label{estofour}
 \dot \phi^{'}\,=\,0.\ee
                                                                                 
Eq. (\ref{estofour}) can be  easily  integrated  to give
\be
\label{estofive}
\phi(t,r)\,=\,\phi_0(r)\,+\,\phi_1(t),\ee                                                                                                        where  $\phi_0$ and $\phi_1$  are arbitrary functions of $r$ and $t$, respectively. Finally, by using (\ref{estotwo}) and (\ref{estofive}), the Euler eq. (\ref{eninetyfour}) can be integrated to give
\be
\label{estosix}
\phi_0(r)\,=\,\frac{1}{2} \log(1-v^2(r)).
\ee                                                                                                                                                                            

\subsection{Weak field limit for the Tolman-Oppenheimer-Volkoff (TOV) equation}
As a weak field limit we understand a space-time described by a small perturbation of the Minkowski metric at sub-Hubble scales (cp. \cite{rnewd}).

To be specific we follow the procedure of Green and Wald \cite{rfifteen} and put ($\epsilon\ll1$) 
   \be
 \label{estoseven}\phi=O(\epsilon),\quad  \lambda=O(\epsilon),\quad  v=O(\epsilon^{\frac{1}{2}}),\quad                    
\partial_r\,=\,O(\epsilon^{-1}),\quad   \partial_t\,=\,O(\epsilon^{-{\frac{1}{2}}}). 
\ee  
                                                                                                                                                                   
Then  we obtain  in leading order:
\begin{itemize}

\item From (\ref{eninetynine}) \be
\label{estoeight}
\rho\,\phi_0'\,-\,2\frac{\rho v^2}{r}\,-\,(\rho v^2)'\,=\,0.
\ee
\item From the 2nd, resp. 3rd Einstein eq. (\ref{eeightynine}) resp. (\ref{eninety}) 
  \be
\label{estonine}
\lambda(r)\,=\,r\phi_0'(r),
\ee                                                                                                                                                                          
which, when combined with the 1st Einstein eq. (\ref{eeightyeight}), leads to the Poisson eq.
\be
\label{estoten}
 4\pi G\rho \,=\,\frac{1}{r^2}\,(r^2\phi_0')'.
\ee                                                                                                                                                                           
\item From (\ref{estosix})  
\be
\label{estoeleven} 
\phi_0\,=\,-\frac{1}{2}v^2.
\ee  
\end{itemize}                                                                                                
If we now insert (\ref{estoten}) and (\ref{estoeleven}) into (\ref{estoeight}) we obtain
\be
\label{estotwelve}
3(r^2\phi_0')'\phi_0'\,+\,2\phi_0(r^2\phi_0')'\,=\,0.
\ee                                                                                                                                                                          
 Multiplying (\ref{estotwelve}) by  $(-2\phi_0)^{\frac{1}{2}}$   (integrating factor) we obtain
 \be
\label{estothirteen}
\left((-2\phi_0)^{\frac{3}{2}}\,(r^2\phi_0')\right)'\,=\,0,\ee                                                                                                                                                                          
which, after integration, leads to the following nonlinear ordinary differential equation for the gravitational potential ( $\beta$ = const.)
\be
\label{estofourteen}
(r^2\phi_0')'\,=\,\frac{\beta}{2}(-2\phi_0)^{-\frac{3}{2}},
\ee                                                                                                                                                                     
which was derived in \cite{rtwo} as the stationary solution of the spherically symmetric, nonrelativistic darkon fluid equations.

\subsection{ Modeling halos}
In \cite{rtwo} we used the numerical solutions of (\ref{estofourteen}) to  determine the circular motion of a star in the potential  $\phi_0$             given by the formula (see \cite{rtwentythree})
\be
\label{estofifteen}
\frac{\hat v^2(r)}{r}\,=\,\phi_0'(r),
\ee                                                                                                                                                                            
where  $\hat v$  is the rotational velocity of the star. Thus, if all stars of a galaxy are in circular motion
the graph of   $\hat v$         gives the galactic rotation curve. We recall that the results reported in \cite{rtwo} are in qualitative agreement with observational data.

\section{Final remarks}
In this  paper we have generalized our nonrelativistic darkon fluid model (NDFM), introduced in \cite{rone} and enlarged and reviewed in \cite{rtwo}, to the framework of General Relativity. Our relativistic model contains, as is the case for the NDFM, no free parameters in its Lagrangian. This feature distinguishes our model, to the best of our knowledge, from all other models for dark energy resp. dark matter. 
The relativistic model reproduces, 
at sub-Hubble scales, the cosmological equations derived from the NDFM (section 5) and 
in the weak field limit the nonlinear differential equation satisfied by the gravitational potential for stationary solutions of the NDFM (section 6).
We recall that the NDFM predicts qualitatively correct values of the late time cosmic acceleration as well as the flat behaviour of galactic rotation curves \cite{rone}, \cite{rtwo}.
Note that the derivation of already known results from approximate solutions of the relativistic model has led to new insights resp. physical interpretations:
our nonrelativistic cosmological equations are different from the FLRW model. The cosmic acceleration is not attributed to a negative pressure ({\it e.g}. a positive cosmological                                                                                                                                         constant) but it is due to a dynamically determined negative energy density. This property turns out to be equivalent to a time-dependent spatial curvature. In this our relativistic model is very different from the model of dipolar dark matter and dark energy advocated by Blanchet and Tiec \cite{rtwentyfour}, \cite{rtwentyfive}. These authors consider in \cite{rtwentyfour} a relativistic action which, to some extent, is equivalent to ours but it differs mainly by the addition of an ad hoc internal force depending on the polarization field. This phenomenological internal force mimics a cosmological constant. Thus, their background model is the $\Lambda$CDM model which is completely different from our model.

Finally we note that the comparison of our nonrelativistic model with the $\Lambda$CDM-model with $H(z)$ data 
(see Fig 1. in \cite{rtwo}) suggests that it will be possible to discriminate between these two models only at larger redshifts.    

We have managed to derive the nonrelativistic gravitational potential produced by a galactic halo in our model from a weak field limit of the generalized Tolman-Oppenheimer-Volkoff (TOV) equation. But in contrast to the original application of TOV (hydrostatic equilibrium within a star) we have
derived and applied the generalized TOV to a non-equilibrium situation given by non-comoving coordinates.

The main aim of the present paper was to present a general relativistic version of the NDFM and to look at its approximations which reproduce either the cosmological or the stationary solutions of the NDFM. This we have achieved but we are aware of the fact that further work on the consequences of the relativistic model is called for.

\end{document}